\documentclass[superscriptaddress,amsmath,amssymb]{revtex4-2}
\pdfoutput=1

\usepackage[utf8]{inputenc}
\usepackage{booktabs}
\usepackage{graphicx}       
\usepackage{xcolor}
\definecolor{color1}{rgb}{0.3333,    0.6863,    0.6784}
\definecolor{color2}{rgb}{0.9922,    0.7490,    0.4353}
\definecolor{color3}{rgb}{0.3686,    0.3098,    0.6353}
\definecolor{color4}{rgb}{0.6196,    0.0039,    0.2588}
\definecolor{color5}{rgb}{0.9176,    0.3647,    0.2784}

\usepackage{hyperref}
\hypersetup{
    colorlinks,
    linkcolor = magenta,
    citecolor = blue,
    urlcolor= blue
}

\newcommand{\EE}{E}
\newcommand{\bs}[1]{\boldsymbol{#1}}

\newcommand{\rev}[1]{#1}

\begin{document}

%\title{Foundations of continuous agent-based modeling frameworks for pedestrian dynamics and their implications}% Force line breaks with \\
\title[Near-future projections in continuous agent-based models]{Near-future projections in continuous agent-based models for crowd dynamics: mathematical structures in use and their implications}% Force line breaks with \\

\author{I{\~n}aki ECHEVERR{\'I}A-HUARTE}
\email{iecheverriah@unav.es}

\affiliation{%
 Depto. de Física y Mat. Apl., Facultad de Ciencias, Universidad de Navarra, E-31080 Pamplona, Spain.
}%
\affiliation{%
 Centro de F\'isica Te\'orica e Computacional, Faculdade de Ci\^encias, Universidade de Lisboa, 1749-016 Lisboa, Portugal.
}%

\author{Antonin ROGE}
\affiliation{
\'Ecole Normale Sup\'erieure de Lyon, CNRS, Laboratoire de Physique, 69364 Lyon, France.
}
\affiliation{%
Universite Claude Bernard Lyon 1, CNRS, Institut Lumi{\`e}re Mati{\`e}re, 69100 Villeurbanne, France.
}%
%\altaffiliation{%
%\'Ecole Normale Sup\'erieure Paris-Saclay, 4 Av. des Sciences, 91190 Gif-sur-Yvette, France; \\
%}%

\author{Olivier SIMONIN}
\affiliation{%
CITI lab., INSA de Lyon, Inria,  69621 Villeurbanne, France.
}%

\author{Alexandre NICOLAS}
\email{alexandre.nicolas@cnrs.fr}
\affiliation{%
Universite Claude Bernard Lyon 1, CNRS, Institut Lumi{\`e}re Mati{\`e}re, 69100 Villeurbanne, France.
}%

\date{\today}% It is always \today, today,

\begin{abstract}
This paper addresses the theoretical foundations of pedestrian models for crowd dynamics. While the topic gains momentum, 
current models differ widely in their mathematical structure, even if we only consider continuous agent-based models. To clarify their underpinning, we first lay \rev{the} mathematical foundations \rev{of} the common hierarchical decomposition into strategic, tactical, and operational levels and underline the practical interest in preserving the continuity between the latter two levels by working with a floor field, rather than way-points. Turning to local navigation, we clarify how three archetypical approaches, namely, purely reactive models, anticipatory models based on the idea of times to collision, and game theory, differ in the way they extrapolate trajectories in the near future. We also insist on the oft-overlooked distinction
between processes pertaining to decision-making and physical contact forces. The implications of these differences are illustrated with a comparison of the numerical predictions of these models in the simple scenario of head-on collision avoidance between agents, by varying the walking speed, the reaction times, and the degree of courtesy of the agents, notably.
\end{abstract}

%Use showkeys class option if keyword
          
\maketitle

%\tableofcontents

\section{Introduction}
Many breakthroughs in modern science have hinged on a reshuffling of the mathematical framework in use. For instance, the ability of modern statistics to account for the
regularities of human anatomical features or social events \cite{quetelet1848systeme,quetelet1869homme} can hardly be dissociated from the mathematical handling of errors and uncertainties and the then-emerging theories of probabilities \cite{pearson1894contributions}. Similarly, the theoretical revisitation of the foundations of economics by von Neumann and Morgenstern \cite{von2007theory} has been instrumental for the progress of modern economics.
Transportation science has largely benefited from the advances in these two disciplines.

Presently, within the field of transportation, modeling pedestrian dynamics is a topic that gains more and more traction, due to both its practical relevance for crowd safety \cite{haghani2023roadmap} and flow management, and its theoretical intricacy \cite{maury2018crowds,corbetta2023physics}. But at the same time, little thought has been dedicated to the soundness of its conceptual foundations.
In fact, one almost takes for granted that pedestrian models should retain the formal structure in place in the fields that inspired them.

Thus, models originating from the field of algorithmic robotics (or computer graphics)
typically rely on the notion of velocity obstacles, i.e., the set of all
velocities leading to a collision before a predefined time horizon \cite{van2008interactive,van2011reciprocal,curtis2013pedestrian}, and the idea that the chosen velocity
should not belong to this set. What matters most is to reach a target while avoiding collisions at all costs. In this sense, mathematical proofs are often given to
guarantee the absence of collisions, at least in some regimes \cite{karamouzas2017implicit}.

Shifting the emphasis from global maneuverability to individual choice, economists and econometricians have employed the structure of  discrete-choice models to find which step is optimal \cite{antonini2006discrete} and adequately calibrate their model \cite{robin2009specification}; each agent then chooses to make a step which optimizes a utility function depending on various factors. The idea of optimal steps was also taken up in the optimal-step model \cite{seitz2012natural}, from a more pragmatic standpoint.

By contrast, physicists have propounded  a line of models that
keep the formal structure of Newton's second law 
and handle interactions between pedestrians in the same way as mechanical forces. This is the case for the celebrated social force model 
\cite{helbing1995social} and its countless extensions and variants \cite{Chraibi2010generalized,seer2014validating,chen2018social}. Contacts and collisions between agents are
then possible, particularly at high density, so that these models are frequently used to study evacuations; besides, they 
heavily (arguably, too heavily) rely on these contact forces to reproduce the collective flow of crowds \cite{sticco2020re}.
The equations of motion underlying these models are structurally similar to those used for simple self-propelled particles, such as active Brownian particles \cite{romanczuk2012active}. But
Moussaid et al. insisted on the more heuristic nature that is actually at play in pedestrians' decisions of motion, 
thus regarding the desired velocity (or self-propulsion velocity in the terminology of active matter) as the output of some heuristic rules; their model was successfully tested against experimental data in a broad range of situations \cite{moussaid2011simple}.
Along similar lines, some of us have recently proposed an agent-based model which feeds the output of a decisional layer, consisting of several contributions, into a mechanical equation of motion; the model was validated in an even broader range of situations \cite{echeverria2022anticipating}.

The present paper is \emph{not} aimed at putting forward yet another specific model. Instead, in the continuation of the short report in \cite{echeverria2023revisiting}, it aspires to delve into the mathematical structure of agent-based pedestrian models and clarify their implications on the dynamics that the models predict. We will start from a very broad perspective and revisit in Sec.~\ref{sec:foundations} the hierarchical decomposition into levels of description, showing that strictly adhering to the distinction between the tactical level and the operational one can be limiting for practical purposes. Then, focusing on local navigation, we will propose a delineation of different modeling branches depending on the way agents project their neighbors' trajectories in the near future; we will also stress the difference between the influence of the  (human and built)  environment on walking choices and their mechanical effects in case of contacts. The theoretical discussion will be illustrated using three archetypal models for operational pedestrian dynamics, introduced in Sec.~\ref{sec:models}. Section~\ref{sec:numerical} will expose to what extent their numerical predictions differ, with an emphasis on the simple scenario of head-on collision avoidance between agents under diverse conditions (various reaction times, degrees of courtesy, etc.). We will draw our final conclusions in Sec.~\ref{sec:conclusions}.

\section{A deep dive into the foundations of pedestrian agent-based models}
\label{sec:foundations}
\subsection{Levels of description of pedestrian dynamics}

Pedestrian flows can be probed at different scales. Typically, 
the modeler's task is split into three levels: a strategic level (\emph{where do I want to go and when?}), a tactical level (\emph{what route do I take to reach this goal?}) and an operational level (\emph{how do I interact with the human and built environment locally, en route?}), using the terminology of \cite{hoogendoorn2004pedestrian}.
In this section, we will enquire into this decomposition and see that, depending on how strictly it is enforced and on the extent to which future is anticipated, different branches of models for local navigation arise. With this clarification in mind, we will
question the relevance of a stark distinction between the tactical and operational levels.

It makes no doubt that the foregoing levels of description are coupled to some extent. For instance, should I contemplate going to the sea, my ultimate \emph{strategic} decision may be influenced by my (tactical) knowledge of \emph{approximately} how much it costs to get there, in human, financial, and environmental terms, while the route choice operated at the \emph{tactical} level may hinge on lower-level features. The decomposition may nonetheless be warranted if each level only depends on a \emph{coarse-grained} vision of the lower levels.

To grasp the possible limitations of this reasoning, let us sharpen the definitions in mathematical terms, namely, in space-time $\Omega \times [0,T]$, where $\Omega \subset \mathbb{R}^2$ is the available
geometric space and $[0,T]$ is the time window under consideration. The evolution of a crowd of $N$ agents
is then represented by a set of $N$ trajectories $\bs{r}_i(t)$, for $i=1...N$ and $t \in [0,T]$. 
From this perspective, for each agent $i$, a strategic choice consists in selecting a target region $\mathcal{T}_i$ in spacetime:
\begin{equation}
\text{{\tiny [STRATEGY]} \ Choose }\mathcal{T}_i \subset \Omega \times [0,T]
\label{eq:strat}
\end{equation}

Multiple path routes $\mathcal{R}_i$, i.e., classes of `equivalent' trajectories defined over the time window $[0,T]$, may lead to the target in space, which we assume is reached by time $T$. If one wants routes to be topologically meaniningful, there are but few possible definitions of such classes of equivalence; two trajectories are equivalent if they belong to the same homotopy class, i.e., if they can be smoothly deformed into one another without crossing an obstacle or a no-go area.

The tactical choice comes down to selecting one of these homotopy classes or routes $\mathcal{R}_i$,
that which minimizes a suitably defined generalized cost $\mathcal{C}_i$:
\begin{align}
&
\text{{\tiny [TACTICS]} \ Select}\ \mathcal{R}_i^{\star} \nonumber\\
&{\mathcal{R}_i^{\star}} = \underset{\mathrm{routes}\ \mathcal{R}_i}{\mathrm{arg\,min}}\  \mathcal{C}_i[\mathcal{R}_i]_0^T. \ \ \ 
\label{eq:cost_route_min}
\end{align}

The route cost can be defined \emph{operationally} as the optimum over all trajectories $\bs{r}_i$ belonging to that route,
\begin{equation}
    \mathcal{C}_i[\mathcal{R}_i]_0^T= \min_{\bs{r}_i \in \mathcal{R}_i} C_i[\bs{r}_i]_0^T .
    \label{eq:cost_tactical}
\end{equation}
Finally, only one trajectory $\bs{r}_i^{\star}(t)$ in this class will materialize at the operational level, during navigation,

\begin{align}
&\text{{\tiny [OPERATIONAL]} \ Select } \bs{r}_i^{\star} \in {\mathcal{R}_i^{\star}}\nonumber\\
& \bs{r}_i^{\star} = \underset{\bs{r}_i \in \mathcal{R}_i^{\star}}{\mathrm{arg\,min}}\ C_i[\bs{r}_i]_0^T.
 \label{eq:cost_operational}
\end{align}

\subsection{Selection of an optimal path by rationally bounded agents}
\label{sub:sel_opt_path}

To make the problem more concrete, we write the (still fairly generic) cost function as

\begin{equation}
    C_i[\bs{r}_i]_0^T=  \int_0^T 
    \underset{\text{running cost}}
    {\underbrace{ c_i[\bs{r}_i(t),\bs{r}_{\rev{j\neq i}}(t)] }} dt
    %+ \underset{\text{terminal cost}}{\underbrace{C_i^T[\bs{r}_i(T)]}}
    \label{eq:cost_running}
\end{equation}

\noindent Here, the running cost $c_i$ depends on agent $i$'s trajectory as well as the other agents' trajectories $\bs{r}_{\rev{j\neq i}}=(\bs{r}_{1},...,\bs{r}_{i-1},\bs{r}_{i+1},...,\bs{r}_N)$; the minimizer is the trajectory that is selected operationally. For concreteness, below, we will study simple running costs that just penalize large speeds and the spatial proximity to another agent with a repulsive potential.
Noteworthily, the tactical choice of Eq.~\ref{eq:cost_route_min} is fully entangled with the operational selection of Eq.~\ref{eq:cost_operational}. Adding to this complexity is the dependence of the individual costs $C_i$ on the other agents' trajectories $\bs{r}_{\rev{j\neq i}}$, highlighted in Eq.~\ref{eq:cost_running}. This is why the optimization problem
does not boil down to finding the local minimum of a function, but is a game \cite{von2007theory} between different players, the agents, each intent on optimizing their cost but unable to control the opponents' decisions. In other words, the minimization
should be understood in the sense of a Nash equilibrium \cite{lachapelle2011mean}, i.e., agent $i$'s `optimal' choice is conditioned on the fact that the other agents' trajectories $\bs{r}_j$, $j \neq i$, are also optimal for them.

Not only does this process make the model computationally complex, but one may also question its realism: \rev{pedestrians are unable to accurately foresee the others’ trajectories; at best, they can try to predict them}, so that
$\bs{r}_{\rev{j\neq i}}$ in Eq.~\ref{eq:cost_operational} should be replaced by its prediction $\bs{\tilde{r}}_{\rev{j\neq i}}^{(i)}$ by agent $i$.

Pragmatically, one can assume that at each time $t$ these predictions are accurate for the short-term future $[t,t+\delta t]$, but are blurry beyond $t + \delta t$, so that the agent needs to adjust their plans of motion every so often, based on their estimation of
\begin{align}
    C_i[\bs{r}_i]_t^T &=  C_i[\bs{r}_i]_t^{t+\delta t}  + C_i[\bs{r}_i]_{t+\delta t}^T 
     \label{eq:approx_cost_operational} \\
    &\approx  C_i[\bs{r}_i]_t^{t+\delta t} + \tilde{C}_i^{t+\delta t}\Big(\bs{r}_i(t+\delta t)\Big), \nonumber
\end{align}
where $\tilde{C}_i^{t+\delta t}\Big(\bs{r}_i(t+\delta t)\Big)$ is a `fuzzy estimate' of the future cost beyond $t+\delta t$ (notice the contrast with the exact treatment of the value function in optimal control theory). Assuming that trajectories cannot be planned to any level of detail in the blurry future, $\tilde{C}_i^{t+\delta t}$ will only depend on the position $\bs{r}_i(t+\delta t)$ reached at that time. Note, in anticipation, that if $\delta t$ is small the trajectories can be considered linear, i.e., $\forall j \neq i,\ \bs{r}_j(t+\delta t) \simeq \bs{r}_j(t) + \bs{v}_j(t) \delta t$, so that the minimization of Eq.~\ref{eq:approx_cost_operational} will ultimately reduce to a dynamical equation of motion: at each time $t$, each agent $i$ has to select an optimal velocity $\bs{v}_i^{\star}$.

Quite interestingly, these abstract considerations
surreptitiously led to the substitution of the remaining cost $C_i[\mathcal{R}]_{t+\delta t}^T$ in the equation of motion by a floor field
(also called potential field or distance function \cite{kretz2009pedestrian,dietrich2014gradient,kleinmeier2019vadere}) $\tilde{C}_i^{t+\delta t}( \bs{r})$,
\begin{equation}
    C_i[\mathcal{R}]_{t+\delta t}^T \rightsquigarrow \tilde{C}_i^{t+\delta t}\Big( \bs{r}(t+\delta t)\Big).
\end{equation}
The latter is a real-valued function $\Omega \to \mathbb{R}$ that quantifies how attractive a given location $\bs{r}$ is effectively, for every class of agents. In principle, it should exactly match the cost of the optimal route from that location to the target location over the time window $[t+\delta t, T]$, but in practice this cost may be assessed roughly, notably in view of the shortest-path distance of $\bs{r}$ to the goal or via an Eikonal equation \cite{dietrich2014gradient,kleinmeier2019vadere}, traditionally used in ray tracing algorithms, in which the cost of walking through a given zone plays the role of an index of refraction \cite{echeverria2022anticipating}.
Finer alternatives can e.g. aim to estimate the remaining travel time from $\bs{r}$ to the goal in light of the current density field, hence letting the pedestrian density play a role akin to the refraction index \cite{kretz2009pedestrian}.

\subsubsection*{Alternative approaches to split the tactical and operational layers}
The use and storage of floor fields presents the major advantage of making the tactical-operational connection virtually seamless,  but it is relatively memory-consuming,  say around a few megabytes per agent type for a $100\,\mathrm{m} \times 100\,\mathrm{m}$ space with a $10\,\mathrm{cm}$ resolution. This used to be an issue in the past, but
this is no longer so with any modern computer.

Perhaps due to this reason, historically floor fields were not used in continuous models and a number of them still rely on
intermediate `way-points' \rev{(a hybrid global-local approach)} that are fed into the operational layer by the tactical one \cite{helbing1995social,pelechano2005crowd,curtis2013pedestrian,ikeda2013modeling}. More concretely, routes $\mathcal{R}$ may be defined as paths in the visibility graph of the environment, in which two areas are linked if they are directly connected and mutually visible. They are thus represented by a series of way-points (also known as intermediate goals, subgoals, or midway destinations), which notably go around large obstacles.
Route planning favors the route with the shortest distance or, in a dynamic approach, the one with the shortest travel time based on current densities.

Previous works have already underscored the practical advantage of letting agents follow the gradient of a space-covering map (or floor field) \cite{kretz2009pedestrian,dietrich2014gradient}, rather than guiding them towards the next way-point (by having their desired velocity point to it).
It may nonetheless be useful to expose this advantage in a more \emph{fundamental} way, in the context of the above reasoning. 
Introducing way-points conveniently decouples the tactical route planning from the local navigation of Eq.~\ref{eq:cost_operational}, but this split may 
have tangible fallout whenever agents end up deviating from the initially set route.

\begin{figure*}[h]
\centering
\includegraphics[width=0.8\textwidth]{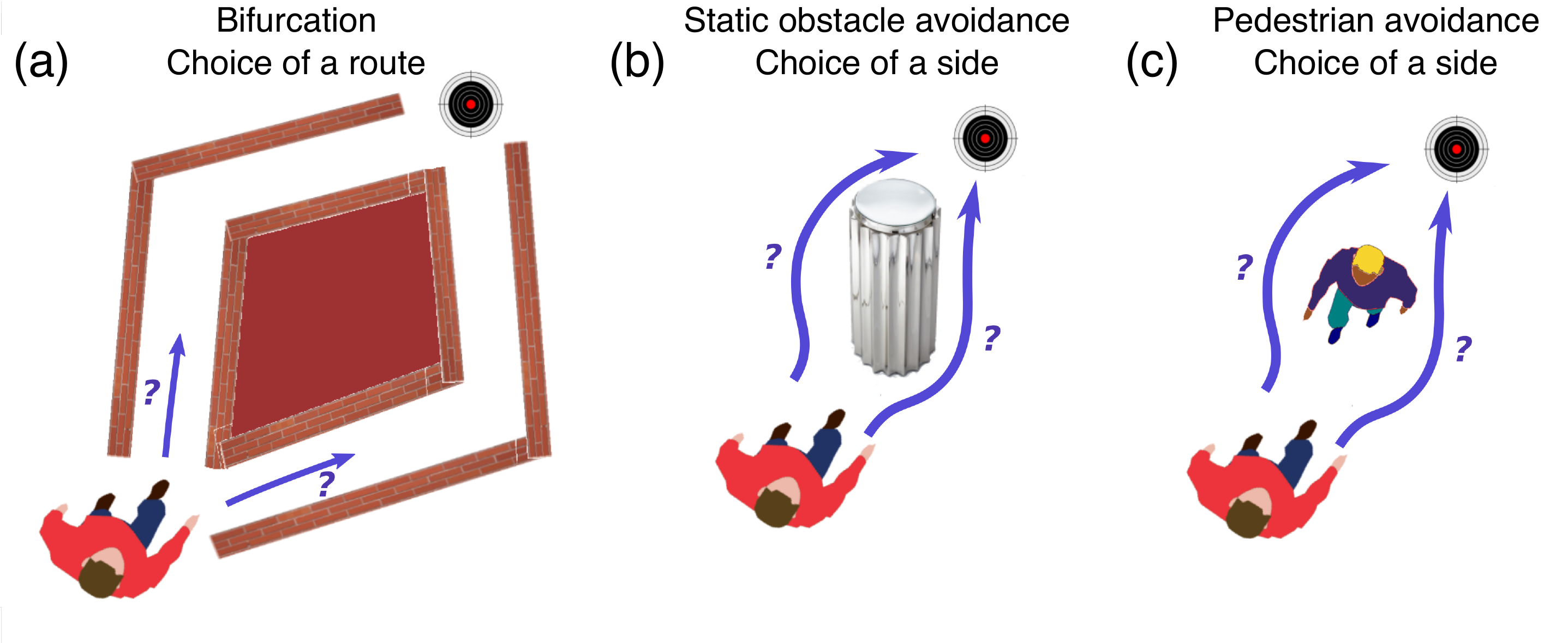}
\caption{Route choice spans a continuum of decisions, from (a) actual routes to (b) obstacle avoidance, or even (c) pedestrian avoidance.}
\label{fig:Bifurcation}       % Give a unique label
\end{figure*}

Consider a situation in which an agent must avoid a large tree on the sidewalk by swerving to the left (which is slightly shorter) or to the right. Adhering to the above decomposition, the route planner will select the left path and set way-points along it. Should a counter-walking 
pedestrian unexpectedly obstruct the left path, the simulated agent will still aim left, whereas in reality they would probably opt for the right path in this circumstance.
In other words, the degeneracy reflecting the quasi-equivalence of two routes at the tactical level was lifted too early. Such problems are expected whenever the accessible space between possible agents' positions and the next way-point becomes non-convex (see the example of Sec.~\ref{sub:obstacle}) or even not simply connected, because of the presence of obstacles. Incidentally, on account of its topological origin (one cannot smoothly deform a route into another one), the problem may occur regardless of the size and nature of the hindrance, that is to say, for extended obstacles that are generally handled by the tactical layer [Fig.~\ref{fig:Bifurcation}(a)] as well as for thin obstacles [Fig.~\ref{fig:Bifurcation}(b)] or even (groups of) pedestrians [Fig.~\ref{fig:Bifurcation}(c)].

\subsection{Modeling local navigation}
\begin{figure*}[h]
\includegraphics[width=\textwidth]{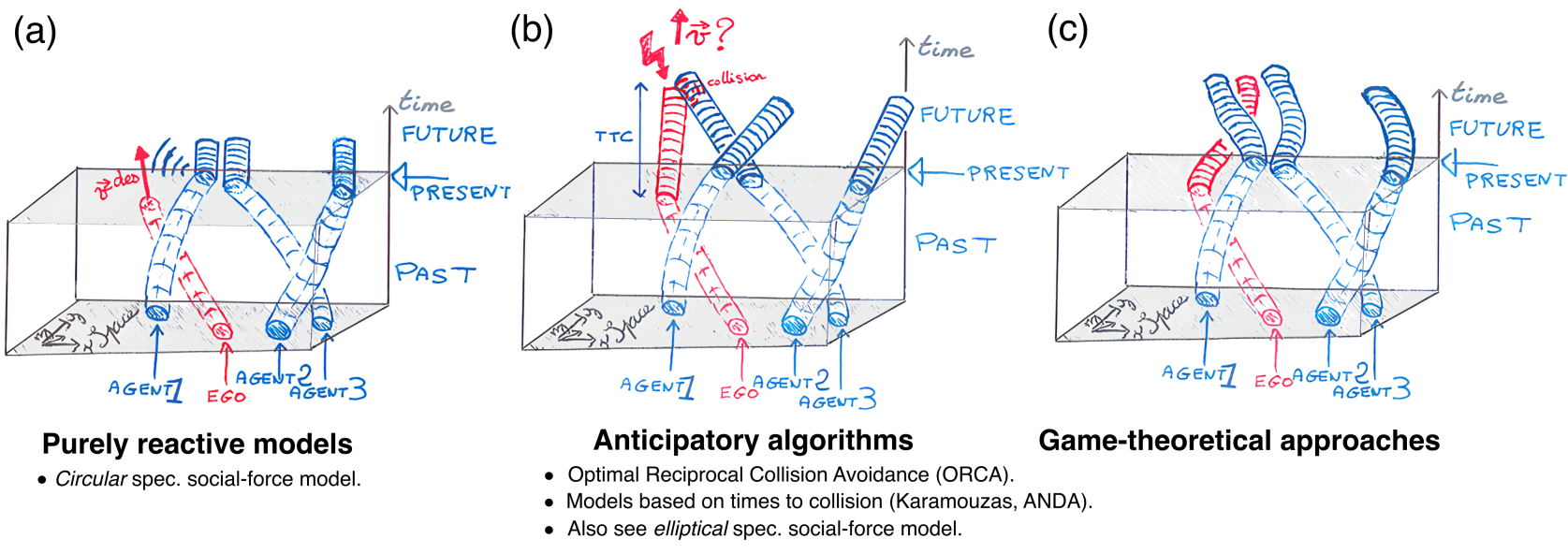}
\caption{Schematic differences between different stems of agent-based models, highlighting their distinct extrapolations of the future. In this spacetime depiction, agents' trajectories are represented as tubes.}
\label{fig:spacetime_sketches}       % Give a unique label
\end{figure*}

No matter how tactical planning is implemented (via a road-map or a floor field as in Eq.~\ref{eq:approx_cost_operational}), 
local pedestrian navigation will also require 
responding to the motion of neighboring pedestrians in the short-term, foreseeable future. Naturally, the reason why these are more difficult to handle than static obstacles is that they are moving. Visualising trajectories as non-overlapping winding tubes (or tunnels) in spacetime (Fig.~\ref{fig:spacetime_sketches}) provides an illuminating perspective on the way the neighbors' motion is handled in different modeling branches.

Purely reactive models treat other agents similarly to static obstacles, i.e., as time-invariant cylinders in spacetime, illustrated in Fig.~\ref{fig:spacetime_sketches}(a). 
Agents thus interact in much the same way as driven passive particles \cite{bain2018hydrodynamics} with a configuration-dependent potential, except that the interactions need not be reciprocal. 
The Social Force Model (SFM) with circular specification \cite{helbing2000simulating} is probably the most well-known model in this branch.

At the other extreme, in a game-theoretical approach, 
in order to steer appropriately, each agent $i$ strives to predict the other agents' motion $\bs{\tilde{r}}_{\rev{j\neq i}}^{(i)}$ in the near future, i.e., furthers their tubes in spacetime; see Fig.~\ref{fig:spacetime_sketches}. Then, they adjust their own tube to optimize the generalized cost
of Eq.~\ref{eq:cost_running}. Importantly, furthering the other agents' trajectories is not merely
a \emph{forecasting} task, insofar as these trajectories depend on agent $i$'s choice of motion; thus, agent $i$ has to anticipate their neighbors' responses.
The numerical resolution of these intricacies requires
 heavy computations (which prompted the development of mean-field versions of this approach \cite{lachapelle2011mean,bonnemain2023pedestrians}).

To keep in check the computational cost, methods based on the anticipated time to collision (TTC) have been put forward and validated empirically \cite{Karamouzas2014universal}.  More precisely, the TTC is computed as the first time in the future at which a collision is expected if the neighbouring agents' velocities are constant. An expression for the TTC-based interaction
energy was derived by Karamouzas and colleagues \cite{Karamouzas2014universal}, in the form of a cut-off power law, under the assumption of circular pedestrian shapes. 
In space-time, basing the potential on the TTC (which is a function of current positions and velocities) comes down to prolonging the other agents' past trajectories as cylinders invariant along their velocity vectors $\bs{v}_j$ [Fig.~\ref{fig:spacetime_sketches}(b)], whereas a full-fledged game-theoretical approach allows any tubular shape extending into the future [Fig.~\ref{fig:spacetime_sketches}(c)].

Depending on whether $\bs{v}_j$ are the neighbors' \emph{observed} velocities or the velocities \emph{expected} by $i$, agent $i$'s task will be either mere forecasting, as in \cite{echeverria2022anticipating}, or anticipation of the neighborhood's response. Optimizing the generalized cost as a function of the whole set of velocities $\{\bs{v}_i\}_{1\leqslant i \leqslant N}$, as in \cite{karamouzas2017implicit},  goes in the latter direction. That being said, the gap between the two options is narrower than it may look, because all agents update their desired velocities at every time step, once they have actually \emph{observed} their neighbors' responses to their choices of motion. Still, we will see in Sec.~\ref{sub:courtesy} that mere forecasting with frequent updates may lead to results very different from anticipation, when studying choices made out of courtesy.

\subsection{Mechanics of locomotion and physical contacts}

Pedestrian motion occupies a singular position among transport systems in that physical contacts can occur in close-to-normal flows at high densities. Thus, in addition to choices, mechanical effects need to be taken into account. In this respect, the motion of pedestrian $i$ (as a physical body of mass $m$ and position $\bs{r}_i $), averaged over a stepping cycle, is governed by Newton's law of motion, as follows,
\begin{equation}
    m \ddot{\bs{r}}_i = m \frac{\bs{u}_i^{\star}-\dot{\bs{r}_i}}{\tau^{\mathrm{mech}}} 
        + \sum_{j\neq i} \bs{F}_{j \to i}^{\mathrm{c}}
        + \sum_{w \in \mathrm{walls}} \bs{F}_{w \to i}^{\mathrm{c}},
    \label{eq:Newton-mech}
\end{equation}
where $\bs{F}_{j \to i}^{\mathrm{c}}$ and $\bs{F}_{w \to i}^{\mathrm{c}}$ denote contact forces exerted by neighbouring pedestrians and walls, respectively. The first term of Eq.~\ref{eq:Newton-mech}, which represents the controllable part of the acceleration \cite{hoogendoorn2003simulation} or
the damped self-propelling force of an active particle,
indicates that the desired velocity $\bs{u}_i^{\star}$ is not reached instantly, but only after characteristic time $\tau^{\mathrm{mech}}$ (a fraction of a second in free space) due to the cyclic human gait or the limited friction with the substrate. Importantly,  $\tau^{\mathrm{mech}}$ only depends on locomotion and mechanical interactions, but on no account on the reaction time. 

To elucidate this distinction between this mechanical relaxation time and the cognitive reaction time, let us consider the example of a pedestrian walking on solid ground versus skating on an ice rink. While the reaction time of the pedestrian in both situations is the same, $\tau^{\mathrm{mech}}$ will be longer on the ice rink due to the more slippery surface; the skater will thus need more time to adjust their movements accordingly. To extrapolate to an even broader context, one may contrast the mechanical response time of a boat with that of a ground vehicle. Notably, $\tau^{\mathrm{mech}}$ is substantially higher for the former, reflecting the specifics of moving on water. These two examples from a broader context highlight the relevance of coupling the mechanical layer centred on Eq.~\ref{eq:Newton-mech} with a decision-making layer determining $\bs{u}_i^{\star}$, which is consistent with the early insight of \cite{hoogendoorn2003simulation}, but quite generally overlooked in practice \cite{helbing2000simulating,moussaid2011simple,Karamouzas2014universal,maury2018crowds}.

\subsubsection*{Alternative approaches}
Indeed, in conventional force-based models \cite{helbing1995social,helbing2000simulating,Karamouzas2014universal}, pseudo-forces are additively inserted into
Eq.~\ref{eq:Newton-mech} to account for the deviations from $\bs{u}_i^{\star}$ due to the local environment (other agents and walls). Conceptually, this is not satisfactory, because it puts these cognition-mediated effects on the same footing as mechanical forces, in particular subjecting them to the same relaxation time scale $\tau^{\mathrm{mech}}$.
It so happens that for walking the cognitive reaction time \footnote{\rev{In reality, this cognitive reaction time amalgamates distinct processes, from the perception of the surroundings to the transmission of motor signals to adjust one's motion.}} $\tau_{\psi}$ is of the same
order of magnitude as $\tau^{\mathrm{mech}}$ and that both (cognitive and mechanical) processes take place within the confines of the same physical entity; the pedestrian. This helps explain the widespread conflation of mechanical and decisional processes. If one considers (instead of a pedestrian) a remote-controlled boat, for which $\tau_{\psi} \ll \tau^{\mathrm{mech}}$ and the control operator and the system are separated in space, the confusion is much more striking.

\section{Presentation of different types of models}
\label{sec:models}

After clarifying the conceptual foundations of major frameworks for agent-based pedestrian modeling and their differences, we aim to illustrate their implications in simulations and, for that purpose, this Section introduces archetypal examples for each of the three lines of models under consideration. 

\subsection{Game-theoretical model}
In Sec.~\ref{sub:sel_opt_path}, we argued that the agent's selection of a path, in interaction with their environment, can be handled as the optimization of a generalized cost function (Eq.~\ref{eq:cost_operational}), subject to the concomitant path choices of other agents, hence a competition between the agents. This defines a `game' in the mathematical sense \cite{maury2021game}. While we have reasoned in very general terms so far, here, to put forward a concrete, minimalistic model, we
make further simplifications: we assume that all agents have the same type of cost function and that it takes the  following simple form

\begin{equation}
\small
C_i\Big( \bs{r}_i | \bs{r}_{\rev{j\neq i}} \Big)  = \int_0^{T} 
\Big[ \lambda\,v_i(t')^2 + \sum_{j\neq i} V( r_{ij}(t') )  \Big] dt' + C_i^T \Big( \bs{r}_i(T) \Big),
\label{eq:specific_cost}
\end{equation}
where the term in the integrand involving the speed $v_i(t')=||\bs{\dot{r}}_{i}(t')||$ penalizes large speeds, $V$ is a simple repulsive distance-based potential, e.g., an exponential function $V(r)= V_0 \exp{\Big(-r/r_c\Big)}$ (using the shorthand $r=||\bs{r}||
$), and 
the terminal cost $C_i^T (\bs{r})= \pm K x$, drives the agent to its goal (here going left or right). 
\rev{In short, an agent $i$, aware of the trajectories $\bs{r}_{\rev{j\neq i}}$ of their neighbors $j$, will opt for a trajectory $\bs{r}_i$ that drives them to their goal (hence the terminal cost $C_i^T$), while not walking excessively fast (hence, the penalty $\propto v_i^2$) and not getting too close to other people (hence the repulsive interactions encoded by $V$ along their path).}

As the \rev{cost $C_i$ can be multiplied by any positive scalar without changing} the position of its minimum, it can be rescaled so that $\lambda=1$. Besides, for an isolated agent, $V=0$, the cost $C_i$ in Eq.~\ref{eq:specific_cost} is minimized for a speed $v=K/2$; therefore, the parameter $K$ can be set to $2 v_d$, where $v_d$ is the agent's preferential speed.

The game-theoretical problem is thus well defined. To find a Nash equilibrium, we solve Eq.~\ref{eq:specific_cost}. 
\rev{
For this purpose, we introduce the value function $U(\bs{r},t)$, i.e., the minimal remaining cost $C_i$ starting from a given position $\bs{r}$ at a given time $t$, and we split it into a short-term part over a very small time window $[t,t+dt]$ and the remaining cost $U(\dots,t+dt)$:
\begin{equation}
    U(\bs{r},t)	=\left[v^{\star}(\bs{r},t)^{2}+\sum_{j\neq i}V\Big(r_{ij}(t)\Big)\right]dt+U\Big(r+v^{\star}(\bs{r},t)dt,t+dt\Big),
    \label{eq:HJB_prel}
\end{equation}
where  $v^{\star}(\bs{r},t)$ denotes the optimal velocity at $(\bs{r},t)$ and thus satisfies  $v^{\star}(\bs{r},t)=\frac{-1}{2}\nabla U(\bs{r},t)$, as can be proven by differentiating $U(\bs{r},t)$,.
After expanding the different terms in Eq.~\ref{eq:HJB_prel} to first order in $dt$, we arrive at the Hamilton-Jacobi-Bellman equation for $U$: 
} 

\begin{equation}
    0 = \frac{\partial U}{\partial t} + \sum_{j\neq i} V( ||\bs{r}_{ij}(t)|| ) -\frac{1}{4} (\nabla U)^2,
    \label{eq:HJB}
\end{equation}
with boundary condition $U(\bs{r},T)= C_i^T ( \bs{r}_i )$. Equation~\ref{eq:HJB} is solved numerically over space and time. At each time step ($dt=0.05\,\mathrm{s}$), assuming that the other agent's trajectory (hence $\bs{r}_{ij}$)  is known, the optimal velocity  is calculated as $\bs{v^{\star}}=-\frac{1}{2} \nabla_{\bs{r}} U$, after solving for $U(\bs{r},t)$ in Eq.~\ref{eq:HJB} using an Euler finite-difference scheme in space and time, and starting from $t=T$; the amplification of numerical instabilities in the computation of derivatives is mitigated by smoothing the $U$-fields with a Gaussian kernel (with standard deviation 0.05~m). 
Supposing that no inertia is at play, the agent updates his or her position with $\bs{r}(t+dt)=\bs{r}(t)+\bs{v^{\star}}\,dt$. 

We then proceed iteratively, starting from initial guesses of the two agents' trajectories $\bs{r}_i(t)$ and $\bs{r}_j(t)$, and then updating $i$ and $j$ alternatively, until convergence. Note that initial guesses help to  reach equilibrium trajectories faster but may constrain their shapes, as several Nash equilibria may exist. On the other hand, we have checked that the algorithm is robust to variations of $dt$ and that the aforementioned rescaling of the cost by $\lambda$ has no impact.

\subsection{ANticipatory Dynamics Algorithm (ANDA) }
\label{sub:ANDA}

% EXPLAIN THAT MODEL INTENSIVELY VALIDATED
% include Table 

% BETTER FRAME THE CONCEPTUAL PB IN THIS

% We aim to show that models designed to circumvent these conceptual issues are practically very effectual. More
% precisely, we will briefly recall the main features of the ANticipated Dynamics Algorithm (ANDA) that we very recently introduced \cite{echeverria2022anticipating}
% based on the foregoing considerations and touch on some of its successes in reproducing crowd dynamics; for a detailed presentation of the model and its results, the reader is referred to \cite{echeverria2022anticipating}.
The above optimization of a full trajectory in relation with the rest of the crowd may become computationally costly for a crowd. To achieve a trade-off between this account of anticipation and numerical tractability, we turn to anticipatory models, and in particular the ANDA algorithm introduced in \cite{echeverria2022anticipating}, which consists of a decision-making layer and a mechanical layer (given by Eq.~\ref{eq:Newton-mech}).
%AN: no physical contact in the following, so leave this aside, applied to disks (a crude approximation of pedestrians' shapes) in two-dimensional space, that undergo frictionless Hertzian interactions if they are in contact with one another or with a wall. 
The desired velocity $\bs{u}_i^{\star}$ 
%\footnote{We use $\bs{u}$ instead of $\bs{v}$ to highlight that the actual velocity $\bs{v}$ may differ from the one chosen by the agent, $\bs{u}$.}  
entering the mechanical layer is obtained from the decision-making layer as the 
velocity that minimizes a cost function $\EE[\bs{u}]$ comprising several contributions (Eq.~\ref{eq:Anda}). We will briefly recall the main features of this function; for a detailed presentation of the model, the reader is referred to \cite{echeverria2022anticipating}.

% Colors remove by Iñaki. "I'm not sure whether colored formulas are accepted in PRE"
% \begin{equation}
% E[\bs{u}] = 
% {E_{biomech}} + 
% {E_{FF}} +
% {E_{inertia}}+ 
% {E_{private-space}}+
% {E_{anticipation}}
% \label{eq:Anda}
% \end{equation}

\begin{equation}
\begin{aligned}
E[\bs{u}] = &\, {E_\mathrm{biomech}} + {E_\mathrm{FF}} + {E_\mathrm{inertia}} \\
& + {E_\mathrm{pers-space}} + {E_\mathrm{anticipation}} \label{eq:Anda}
\end{aligned}
\end{equation}

% \begin{center}
% \includegraphics[width=0.9\textwidth]{Minimizaton.png}    
% \end{center}

\begin{figure}[h]
\centering
\includegraphics[width=6.8cm]{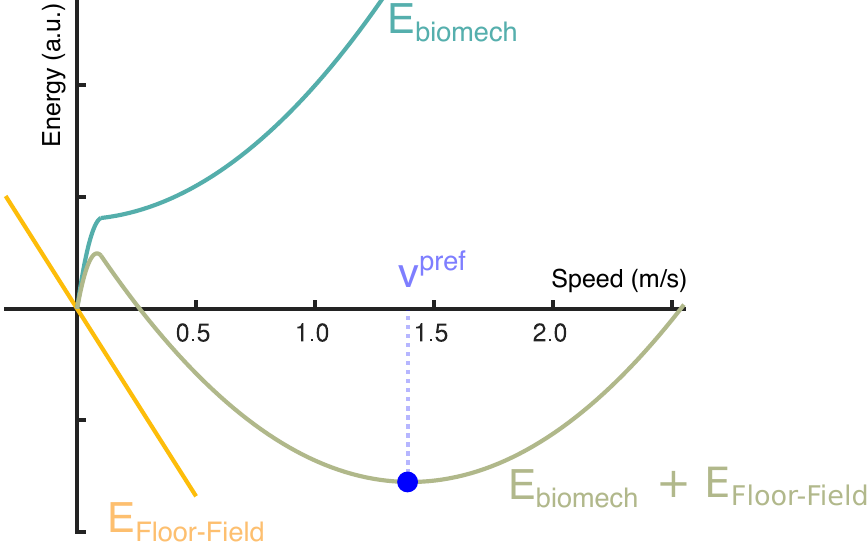}
\caption{Specification of the ANDA model presented in Sec.~\ref{sub:ANDA}. Variations of the bio-mechanical cost $\EE_{\mathrm{biomech}}$ (based on the data from \cite{ludlow2016energy}) and of the floor field $\EE_{\mathrm{FF}}$ with the test speed; by definition, the preferential speed $v^{\mathrm{pref}}$ minimizes the sum of these two contributions.}
\label{fig:V_pref}       % Give a unique label
\end{figure}

\noindent In free space, only three factors are active: the bio-mechanical contribution $\EE_{\mathrm{biomech}}$, which measures the empirical physiological cost of walking at a given speed $u= ||\bs{u}||$ \cite{ludlow2016energy}; the static floor field $\EE_{\mathrm{FF}}$ discussed in Section~2, evaluated at the position to be reached at the next time step with the test velocity $\bs{u}$; and the quadratic penalty $\EE_{\mathrm{inertia}}$ for abrupt changes in velocity.  In uniform motion (i.e., no inertial effect), the chosen velocity is then directed along the gradient of $\EE_{\mathrm{FF}}$ and its magnitude $v^{\mathrm{pref}}$ minimizes the sum of the first two contributions, as illustrated in Fig.~\ref{fig:V_pref}. Accordingly, if one knows an agent's free-walking speed $v^{\mathrm{pref}}$, the slope of the floor field can directly be obtained and the model contains no
adjustable parameter at this point.

On top of these three contributions, for pedestrians walking \emph{alone} (no groups), interactions with the built environment and the crowd generate two new terms, reflecting
two distinct types of repulsive interactions at play in pedestrian dynamics. The first one, $\EE_{\mathrm{pers-space}}$, 
is based on the separation distance between an agent and their neighbours, with a short-ranged repulsive strength decaying with distance, which is familiar to physicists; it reflects the
desire of people to preserve a personal space around themselves, whose extent may vary between individuals and between cultures (as studied by the field of proxemics \cite{hall1969hidden}). Beyond these concerns for personal space, pedestrians also pay particular attention
to the risk of future collisions and adapt their trajectories to avoid them. Karamouzas et al. demonstrated, using empirical data sets, that these effects are much more readily described using a new variable, the anticipated time to collision (TTC), than distances \cite{Karamouzas2014universal}. 

In the ANDA model, we have kept this TTC-based
energy, $\EE_{\mathrm{anticipation}}$, except that non-physical collisions between personal spaces are also taken into account (which results in a smoother profile) and only the most imminent collision is considered.

\begin{figure*}[h]
\includegraphics[width=0.95\textwidth]{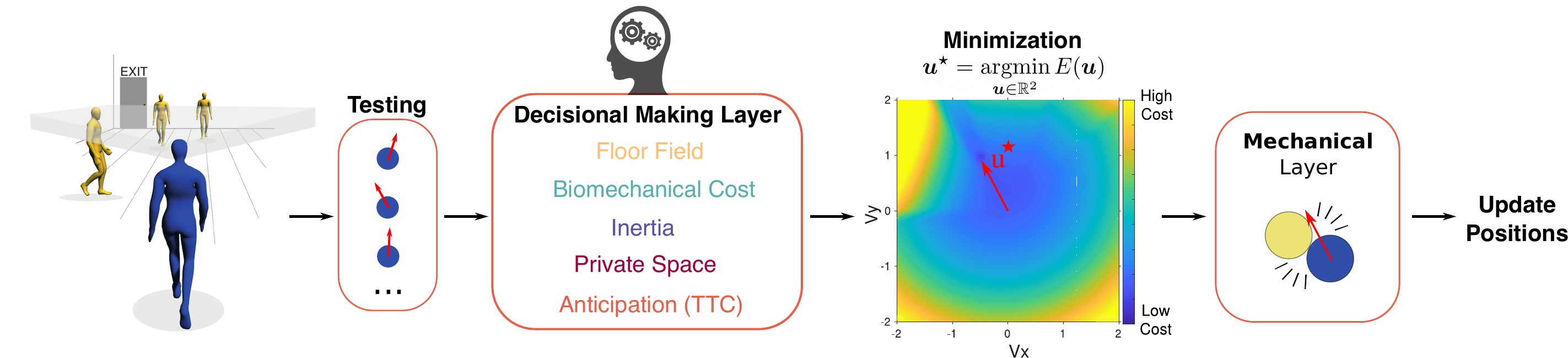}
\caption{General functional diagram of the ANDA model introduced in Sec.~\ref{sub:ANDA}.}
\label{fig:Sketch_model}       % Give a unique label
\end{figure*}

This search for an optimal velocity $\bs{u}_i^{\star}$ minimizing $\EE[\bs{u}_i]$ is done every $\tau_{\psi}$ seconds, which corresponds to the aforementioned \emph{cognitive reaction} time. The actual velocity is then computed using the mechanical equation of motion,  Eq.~\ref{eq:Newton-mech}.
In particular, it may happen that the selected desired velocity $\bs{u}_i^{\star}$ leads to a collision within $\tau_{\psi}$ and thus activates repulsive mechanical forces, an issue which is already addressed in \cite{echeverria2022anticipating}, but into which we shall not delve here, because the examples provided below do not involve any physical contacts.

Incidentally, the cost $\EE[\bs{u}_i]$ involves not only the test velocity $\bs{u}_i$ of agent $i$, but also the velocities $\bs{u}_j$ of the other agents, in particular $i$'s neighbours. Here, we posit that the velocities to take into account for these other agents are the current ones, i.e., $\bs{u}_j=\bs{v}_j(t)$, which comes down to assuming that agent $i$ furthers the other agents' trajectories on the basis of the velocities that he or she currently observes. This assumption departs from what is typically done to determine equilibrium points in game theory, where each agent considers a situation in which the other agents' choices are also optimal for them. Nevertheless, to lowest order in
$\bs{\delta u}_j=\bs{u}_j(t+\delta t)-\bs{u}_j(t)$, the difference is transparent, insofar as a first-order expansion of $\EE[\bs{u}_i]$ only involves terms in $\bs{\delta u}_i$ and in $\bs{\delta u}_j,\,j\neq i$ (no cross terms), and the latter do not affect the minimizer's value $\bs{u}_i^{\star}$.

Overall, the model follows the functional diagram outlined in Fig.~\ref{fig:Sketch_model}. There
are only a few parameters (4 to 6, depending on how they are counted) that can be freely adjusted, including  the spatial extent and the strength of the repulsion from the personal space and the penalty for abrupt velocity changes.
The predictions of the model have already been validated
in a very wide range of scenarios, listed in Table~\ref{tab:Scenarios}, generally in a quantitative way \cite{echeverria2022anticipating}.

\begin{table*}[h]
\caption{Overview of empirical and experimental data used to validate the ANDA model.}
\centering
\begin{tabular}{@{}lll@{}}
\toprule
\multicolumn{1}{c}{\textbf{Description}}      & \textbf{Type}                  & \multicolumn{1}{c}{\textbf{Reference}} \\ \midrule
Avoidance maneuvers of individual pedestrians & Empirical \& Experimental Data              & \cite{Moussaid2009experimental, corbetta2018physics}              \\
Intruder                                      & Experimental Data              & \cite{nicolas2019mechanical,bonnemain2023pedestrians}              \\
Speed density relation - Unidirectional Flow  & Empirical \& Experimental Data & \cite{Older1968,Mori,Weidmann1993,zhang2011transitions}              \\
Speed density relation - Bidirectional Flow   & Experimental Data              & \cite{zhang2012ordering}              \\
Lane Formation                                & Experimental Data              & \cite{jin2019observational}              \\
Bottleneck Flow                               & Experimental Data              & \cite{predtechenskii1978planning,kretz2006experimental,seyfried2009new}              \\
Smartphone Distraction                             & Experimental Data               & \cite{murakami2022spontaneous}              \\ \bottomrule
\end{tabular}
\label{tab:Scenarios}
\end{table*}

\subsection{Social-force model with \emph{circular} specification}
Finally, going all the way down the scale of complexity, we consider a reactive agent-based model.
A classic paradigm of this category of models, the Social-Force-Model (SFM) \emph{in its minimal version}, hypothesizes that the local rules of navigation in a crowd system can be formalized by only using a mechanical layer identical to Eq.~\ref{eq:Newton-mech}, which combines three different forces \cite{helbing2000simulating}. Formally, a pedestrian $i$ who wants to move in a particular direction $\hat{e}_i$ at a desired speed $v_d$, is attracted to this destination by a driving force $\bs{f}^D_i$ which describes the adaptation of his/her current velocity $\dot{r}_i$ to his/her desired one as:

    \begin{equation}
    \bs{f}_i^D = \frac{v_d\,\hat{e}_i-\dot{\bs{r}_i}}{\tau} 
    \label{eq:HelbingDrivinfForce}
    \end{equation}
where $\tau$ is the time needed for the velocity adjustment ($\tau=0.4\,\mathrm{s}$ below). Importantly, in this expression both $v_d$ and $\tau$ are parameters that remain constant over time, regardless of the conditions in which the pedestrian is found. Herein lies the main difference with the ANDA model, where, as explained above, the decisional layer provides the optimal value for the desired velocity by balancing several contributions. 

As pedestrian $i$ moves through space, s/he is repelled by other pedestrians $j$ under the effect of a social force $\bs{f}^S_{ij}=-\bs{\nabla} V_{ij}(r_{ij})$ that mimics the interpersonal distances desired by people when walking. Thus, the ideal path that an isolated pedestrian would follow is permanently modulated by his/her tendency to move away from other individuals. For this purpose, here we use a repulsive function decaying with distance, $V_{ij}(r_{ij})=V_0 \exp(-r_{ij}/r_c)$, following \cite{helbing2000simulating}. This version of the potential, known as the \emph{circular} specification, is the simplest, as it depends solely on the relative distance between pedestrians. (Note that a more sophisticated potential -- the \emph{elliptical} specification -- had been proposed in the paper which originally introduced the SFM \cite{helbing1995social} and accounts for some degree of anticipation of imminent collisions \cite{garcia2023limited,hu2023anticipation}). Finally, physical contacts between agents might happen. To prevent pedestrians from overlapping either with each other or with walls, contact forces $\bs{f}^C_{ij}$ are introduced (but of little use for what follows). 
%Note In their original version, the expressions for these forces were inspired by granular interactions, being composed of a normal elastic and a tangential dissipative terms.

Thus, at each instant $t$, the acceleration of a pedestrian is given by the sum of the internal and external forces to which they are subjected, leading to an evolution of their speed as:

\begin{equation}
m_i \, \bs{\ddot{r}}_i = \bs{f}^D_i + \sum_{j\neq i} \bs{f}^S_{ij}(t) + \sum_{j\neq i} \bs{f}^C_{ij}(t)
\label{eq:Helbingupdate}
\end{equation}
This ordinary differential equation is solved numerically with time step $dt=0.05\,\mathrm{s}$.

\section{Numerical comparison of the output in specific settings}
\label{sec:numerical}
We will now make use of numerical simulations of the paradigmatic models exposed in the previous section to compare their output. In doing so, our primary goal is not
to rank the performances of the models, but to highlight how the \emph{conceptual} discrepancies between modeling branches impact the predictions in concrete settings.
The scenarios under study could be multiplied \emph{ad infinitum}; here, we put an emphasis on two simple, but ubiquitous tasks of binary collision avoidance, amenable to intuitive interpretation: obstacle avoidance and head-to-head collision between two pedestrians

\subsection{Local navigation around complex static obstacles}
\label{sub:obstacle}
First, consider the avoidance of a static obstacle. This example will notably illustrate the above claim that splitting the tactical and operational layers by introducing way-points may drastically fail in some circumstances. Recall that, when the modeler resorts to way-points, the avoidance of small obstacles that do not lie on the segment joining successive way-points is often deferred to the operational module.

This failure is conspicuous in Fig.~\ref{fig:static_obs_avoidance}(a,c), where the simulated agent starts behind a square obstacle or a non-convex obstacle, supposedly reached as a consequence of interactions within the crowd, and then finds itself entrapped.

Technically speaking, the obstacle was made of adjacent columns, each exerting the same repulsion as a static agent, $| V_{ij}^{S} | = V_0 \exp(-r/r_c)$, with \rev{$V_0=100$} and \rev{$r_c=0.75\,\mathrm{m}$}; the fairly large value of $r_c$, which implies that repulsion extends over several meters, is necessary to prevent collisions when the desired speed is increased. We observe that the agent [simulated with the circular SFM in Fig.~\ref{fig:static_obs_avoidance}(a,c)] is blocked ahead of the obstacle. Lowering the repulsion to  \rev{$V_0=20$} is not conducive to a more realistic response for concave obstacles: the agent then succeeds in entering the concave shape but gets trapped in it [Fig.~\ref{fig:static_obs_avoidance}(c)]. (In both types of simulations, a small noise was introduced on the initial positions to break the symmetry along the $y$-axis.)
Other agent-based models in which desired velocities are prescribed independently of the obstacles would fail similarly \cite{echeverria2022anticipating}.

In contrast, combining these operational models with a dynamic potential or floor field determined throughout space sweeps away the tactical/operational divide and overcomes the foregoing entrapment issue.
Within ANDA, the extensive information about the geometry contained in the floor field $V(\bs{r})$ naturally guides the agent around the obstacle, irrespective of its shape [Fig.~\ref{fig:static_obs_avoidance}(b,d)]. 
\rev{In fact, even in the circular SFM, substituting the fixed desired direction with the gradient of this floor field yields a hybrid SFM that reproduces similar avoidance behavior (see the supplementary material and Fig.~S1). Analogous ideas were implemented in several prominent SFM variants \cite{kretz2009pedestrian,kretz2011quickest}.}
In game-theoretical approaches, this geometric information is contained in the time-dependent value function $u(\bs{r},t)$.

\begin{figure*}[h]
\includegraphics[width=0.8\textwidth]{Obst_Avoidance.pdf}
\caption{Avoidance of a static obstacle of convex square shape or non-convex shape. (a and c) Prediction of a way-point-based circular SFM for an agent walking at $v_d=1.5\,\mathrm{m/s}$ towards a way-point located behind the obstacle; the obstacle is made of adjacent columns, each generating a potential  (a) with high repulsion strength \rev{$V_0=100$} or (c) low strength \rev{($V_0=20$)} ; (b and d) Predictions of ANDA, for which the floor field integrates the effect of the obstacle.}
\label{fig:static_obs_avoidance}       % Give a unique label
\end{figure*}

\subsection{Frontal collision avoidance}
We now turn to  collision avoidance between two pedestrians, with a particular interest in the qualitative features obtained when the desired speed is varied from slowly walking agents to people running towards one another. To simulate this situation, we consider a scenario similar to the experiments of Moussa\"id et al. \cite{Moussaid2009experimental}, with two agents initially positioned at opposite ends of a corridor, about 10 meters away from one another and intent on reaching the other end of the corridor. Initially, both agents are slightly displaced in the vertical direction following a uniform random distribution centered at 0, with an amplitude of \(r/2\), where \(r\) is the radius of the pedestrian. This ensures that they will always collide assuming straight trajectories. A total of 100 repetitions were simulated with varying initial conditions. Note that, in the three models under consideration, the increase in the desired speed reflects higher eagerness or hurry to reach the target, and not a leisurely jogging session.

\begin{figure*}[h]
\includegraphics[width=0.8\textwidth]{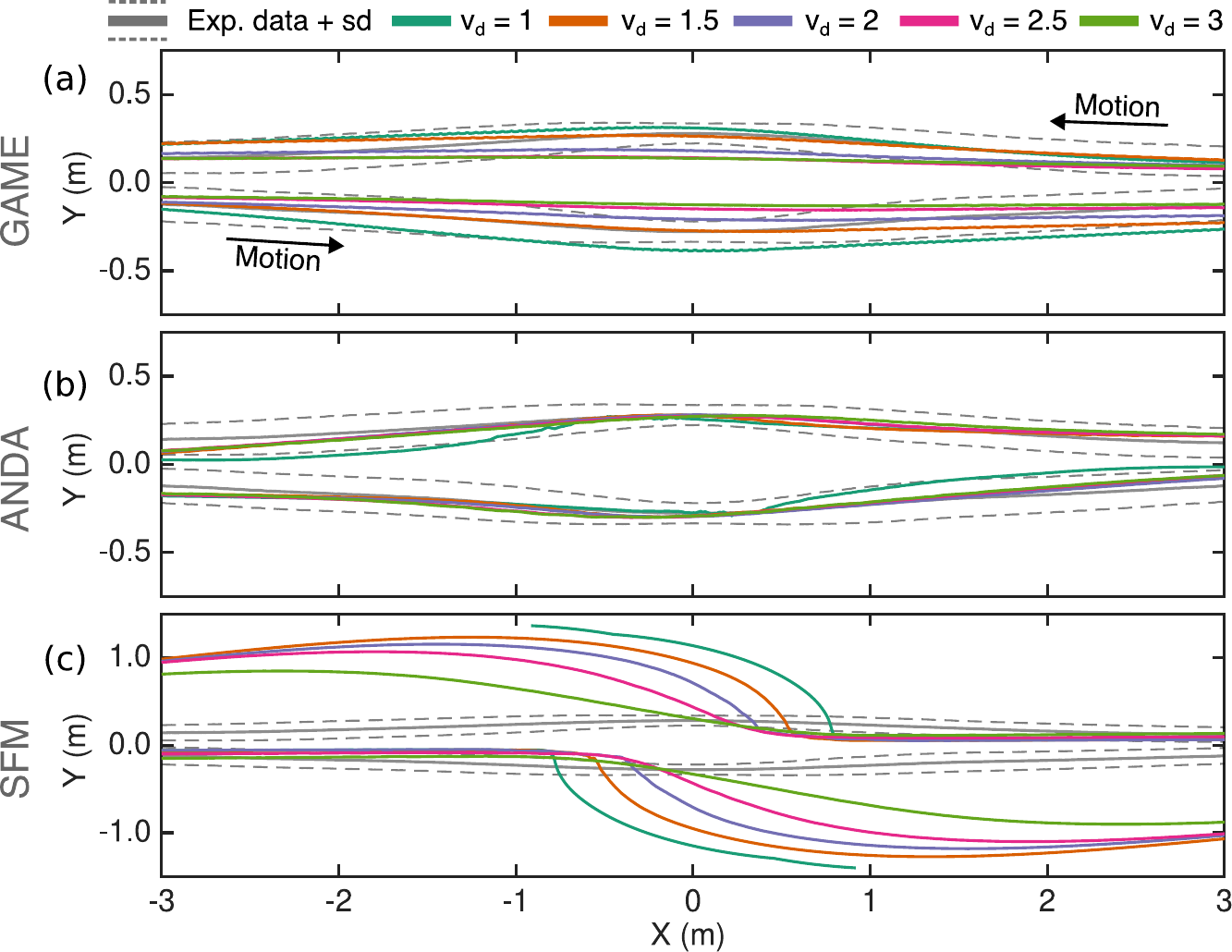}
\caption{Collision avoidance by two counter-walking agents in a straight corridor, as a function of the preferential speed $v_d$. Numerical predictions based on (a) game theory, (b) ANDA, (c) SFM.  The experimental data, drawn in gray, come from \cite{Moussaid2009experimental}.}
\label{fig:CollAvoid}       % Give a unique label
\end{figure*}

It has been shown that an SFM with forces spatially tailored in an \emph{ad hoc} way can replicate the experimentally obtained mean trajectories for \rev{leisurely} walking pedestrians \cite{Moussaid2009experimental}. These trajectories are also quantitatively reproduced by the ANDA model \cite{echeverria2022anticipating}, as shown in Fig.~\ref{fig:CollAvoid}(b). 

With the simple SFM introduced above, without specific tailoring of the potential $V$, the replication of the experiments is of course much poorer [Fig.~\ref{fig:CollAvoid}(c)]. For most parameters that have been tested, collisions are observed when the desired speed is increased. The only way to prevent a collision between the agents at all desired speeds is to impose a strong repulsion, \rev{$V_0=100$}. Then, at moderate speeds, e.g., $v_d=1.5\mathrm{m/s}$, this leads to a sharp and strong detouring behavior, represented in Fig.~\ref{fig:CollAvoid}(c). More interesting are however the qualitative changes that occur when $v_d$ is further increased. As the interaction is only based on distance, the avoidance maneuver is then undertaken when collision is really imminent, i.e., much later than one would expect. Therefore, despite the repulsion strength, at $v_d=3\mathrm{m/s}$ the agents even fail to avoid one another, even though they started 10 meters away from each other.

The situation differs widely with ANDA: since short TTC  (and not only short distances) are heavily penalized in the selection of an optimal velocity, the agents will start interacting farther and farther ahead as $v_d$ is increased, reflecting anticipation of the upcoming colllision. As a matter of fact, the spatial profiles of the trajectories in Fig.~\ref{fig:CollAvoid}(b) do not change much when $v_d$ is varied: at higher $v_d$, at a given distance ahead of the collision point, the stronger TTC effect tends to be balanced by the higher eagerness to move forward. In any event, the agents manage to avoid collision in all these circumstances. \rev{Notably, a minor modification of the baseline SFM—substituting the \emph{circular} specification with an \emph{elliptical} formulation that incorporates a degree of anticipation—leads to markedly improved performance (see Supplementary Information, notably Fig.~S2). This result emphasizes the critical role of anticipatory interactions in effective collision avoidance.}

This also holds for the game-theoretical model, where the effect of varying $v_d$ is however felt slightly more strongly [Fig.~\ref{fig:CollAvoid}(a)]. This is due to the proportional increase of the cost driving the agent to its goal, which starts to dominate the repulsion with other agents. Technically speaking, here we considered a repulsive strength $V_0 = 1.1$,  which provides acceptable results with respect to the empirical data. This parameter was chosen so as to minimize the distance between the experimental trajectory and the simulated one, measured in terms of the cumulative error between each simulated trajectory point the closest point of the real trajectory, and to reach an inter-agent distance when passing as close as possible to the experiments. Besides, note that we have added to the terminal cost $C^T$ a term $K_y\,|y|$ (with $K_y = 1.5$), which pulls agents towards the central line, in order to limit their diffusion along the $y$-axis.

\rev{\subsection{Characteristic Times and Their Effects: Distraction and Mechanical Friction}}

Let us now underscore the differences between the various timescales characterising pedestrian motion and appearing distinctively in ANDA. 
 
\rev{The first timescale is the \emph{cognitive reaction time} $\tau_\psi$. It encompasses both how often the agents refresh their perceptions of the environment (gazing activity) and the short delay before they adjust their motion in response to their observations. In the ANDA model, $\tau_\psi$ corresponds to the interval} between updates of the desired velocity given by the decisional layer. Obviously, this time is expected to soar if the person is engaged in a discussion or playing with their smartphones (texting or web-browsing, in particular) \cite{ropaka2020investigation,murakami2021mutual}. 
 
\begin{figure*}[h]
\includegraphics[width=0.8\textwidth]{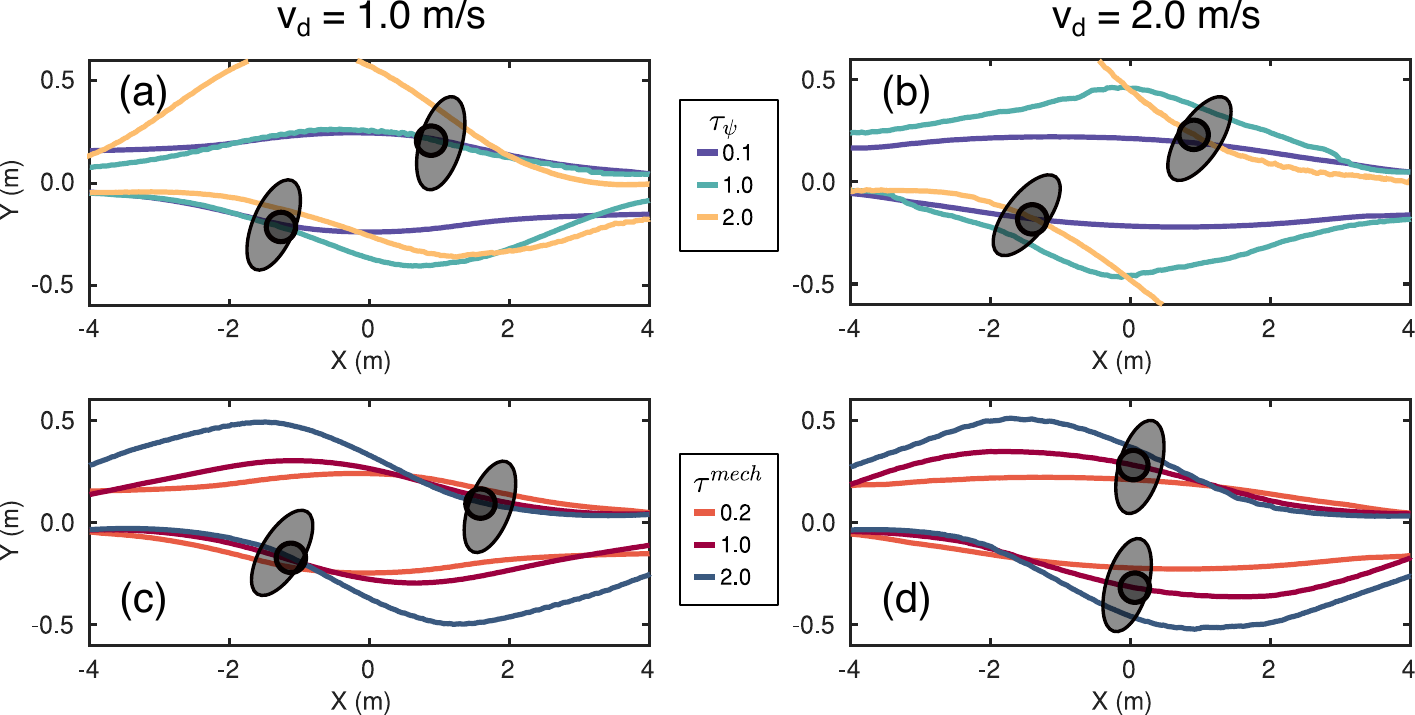}
\caption{Effect of the cognitive reaction time $\tau_{\psi}$ and the mechanical relaxation time $\tau^{\mathrm{mech}}$ on the collision avoidance between two agents in a straight corridor \rev{using the ANDA model}. First row (varying time-update) for $v_d$ = 1.0~m/s (a) and $v_d$ = 2.0~m/s (b). Second row (varying $\tau^{\mathrm{mech}}$) for $v_d$ = 1.0~m/s (c) and $v_d$ = 2.0~m/s (d).}
\label{fig:CollAvoid_times}       % Give a unique label
\end{figure*}

The consequences of this are illustrated in Fig.~\ref{fig:CollAvoid_times} for binary collision avoidance. At low or moderate walking speed $v_d=1\,\mathrm{m/s}$, 
a delayed reaction (larger $\tau_{\psi}$) implies that the agents persist longer on their initial, collisional paths (especially when $\tau_{\psi}$ increases from $1\,\mathrm{s}$ to $2\,\mathrm{s}$), but then swerve more abruptly and more markedly. These trends are even more pronounced at higher walking speed, $v_d=2\,\mathrm{m/s}$. The results compare qualitatively very well with the  experiments conducted by Murakami and colleagues \cite{murakami2022spontaneous}. In these experiments, a series of binary collision avoidance maneuvers was performed, in which one of the two pedestrians was sometimes asked to perform some complex activity on their smartphone while passing their counterpart. Interestingly, compared to the baseline with no smartphone distraction, the avoidance maneuver by a distracted agent is undertaken later and is more abrupt
(see Fig.~1B and S1 of \cite{murakami2022spontaneous}). Meanwhile, in the experiments, the non-distracted agent, who could not fully rely on mutual coordination for this avoidance, undertook a somewhat larger detour. The distance when passing the distracted agent is ultimately larger on average than between two non-distracted participants, by 5 to 10~cm. The increase of the distance when passing as agents are more distracted (i.e., higher $\tau_{\psi}$) is also observed in ANDA (Fig.~\ref{fig:CollAvoid_times}): By responding with sufficient anticipation, gradual and limited adjustments of the velocity are sufficient, whereas abrupt detours may be needed if a collision is perceived only when it is imminent. 
Nevertheless, since $\tau_{\psi}$ cannot be obtained from a quantity measured in the experiments and ANDA probably does not capture all subtleties associated with digital distraction, we cannot provide a more quantitative comparison. 
  
The second timescale is the  mechanical relaxation time $\tau^{\mathrm{mech}}$. Large $\tau^{\mathrm{mech}}$ denote a more inertial response, which would be typical of ice-skaters or swimmers, hence a difficulty to hold or recover one's course if the presence of inertia is not internalised enough (bottom of Fig.~\ref{fig:CollAvoid_times}). This echoes the advice given to sailors to steer the wheel smoothly and with anticipation.
 
\rev{It is worth adding a note on the \emph{numerical integration time} $dt$ in continuous models. Typically much shorter than both $\tau_\psi$ and $\tau^{\mathrm{mech}}$, this time step is a computational artifact, chosen to obtain stable and accurate numerical scheme. It must be clearly distinguished from the cognitive and physical  timescales, $\tau_\psi$ or $\tau^{\mathrm{mech}}$, insofar as it does not reflect any real delay or inertia in pedestrian behavior and has no genuine physical counterpart.}

\subsection{Priority and courtesy}
\label{sub:courtesy}
Finally, one last example will showcase the relevance of establishing formal connections between the different modeling frameworks in Sec.~\ref{sec:foundations}. It deals with
the way in which tacit cultural and behavioral codes, dubbed `intangible factors' in \cite{curtis2013pedestrian}, can be accounted for. Concretely, one is concerned with the priority given to some people, e.g. out of courtesy, when walking. Such effects were rendered by means of \emph{ad-hoc} proxy agents in \cite{curtis2013pedestrian}, that is, specifically devised virtual `outgrowths' of an agent's body that neighbours are forced to heed when moving.

We argue that these effects can be described  much more naturally and generally thanks to the parallels exposed in Sec.~\ref{sec:foundations}. Courtesy, priority, and possibly other behavioral rules can be rendered by internalizing a fraction $\alpha>0$ of the cost (or utility) experienced
by the higher-priority agent ($j$) in the `courteous' agent's cost $C_i$, that is, modifying the game-theoretical cost functions entering Eqs.~\ref{eq:cost_route_min}-\ref{eq:cost_operational} as follows:
\begin{equation}
    C_i[\dots] \to (1-\alpha)\,C_i[\dots] + \alpha C_j[\dots].
\end{equation} 
This is an example of \emph{altruistic} preferences. Incidentally, as emphasized elsewhere in a more general context \cite{alger2013homo}, this altruism does not result from stiff (Kantian) morality: the \emph{homo kantiensis} of Alger and Weibull  \cite{alger2013homo}, who would strive for the best outcome in would-be encounters with their likes, would not exhibit any form of such courtesy; on the other hand, note the conceptual similarity with the methods put forward by Hoogendoorn et al. in a different context \cite{hoogendoorn2023game}.
Naturally, these ideas can straightforwardly be extended to situations with several higher-priority neighbours $j$.

Now, our success in framing the equations of motion of anticipatory models (such as ANDA) as simplified games in Sec.~\ref{sec:foundations} straightforwardly paves the way for translating
 costs in game theory into new pseudo-energies $\tilde{E}_i$ in ANDA (see Eq.~\ref{eq:Anda}). The decisional layer of agent $i$ will then consist in 
minimizing $\tilde{E}_i[\bs{u}_i]$, where 

\begin{equation}
    \tilde{E}_i[\bs{u}_i] = (1-\alpha)\ E_i[\bs{u}_i  \, |\,  \bs{v}_j(t)] + \alpha \ \underset{\bs{u}_j \in \mathbb{R}^2}{\min}\,E_j[\bs{u}_j \, |\, \bs{u}_i].
\end{equation}

\begin{figure}[h]
\centering
\includegraphics[width=10cm]{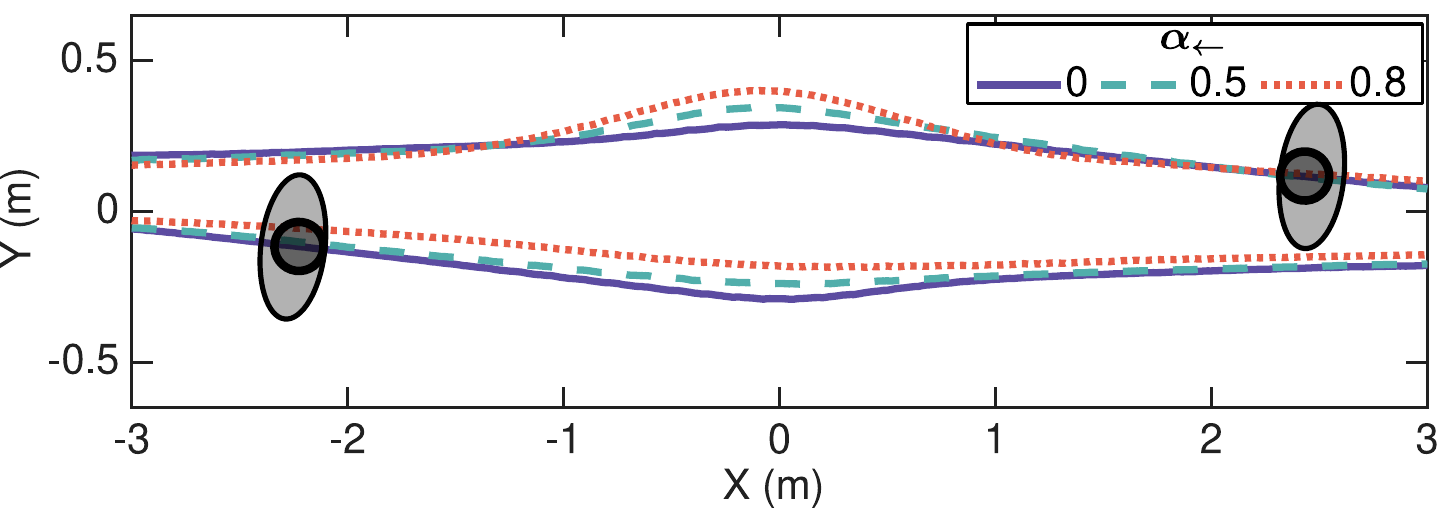}
\caption{Effect of courtesy \rev{in the ANDA model}: Binary collision avoidance between a standard agent moving from left to right ($\alpha_{\rightarrow}=0$) and a courteous agent moving from right to left with different degrees of courtesy $\alpha_{\leftarrow}$, as specified in the legend.}
\label{fig:Courtesy}       % Give a unique label
\end{figure}

To illustrate the outcome of this element, we turn back to our example setup and simulate the head-on collision avoidance between a courteous pedestrian ($\alpha>0$) and an elderly person with walking issues ($\alpha=0$) and show the result in Fig.~\ref{fig:Courtesy}. Clearly, with increasing courtesy (or altruism) $\alpha$, the courteous agent (willingly) takes a larger and larger share of the effort required to avoid a collision. Qualitatively, this is in line with the empirical results of collision avoidance between a young adult and an old one reported in \cite{rapos2021collision}: the young adult contributes a larger share of the avoidance maneuver than the (supposedly, `higher-priority') old one. (In the experiments, the distance when passing was also greater in that situation, as compared to the avoidance of two young adults, which is not replicated in our simple model.)

Interestingly, courtesy could not have been accounted for by simply enhancing the 
agents' \emph{forecasting} abilities in ANDA (or any similar model), e.g.~by letting them prolong the others' tubes in spacetime on the basis of not only the observed velocities, but also the observed accelerations. Instead, the courteous move hinges on the anticipation of how the other agent might move \emph{in response to one's own hypothetical actions}. Similarly, a purely selfish strategy, whereby one undertakes no collision avoidance whatsoever, is only viable if one anticipates that the other agent will swerve.

\section{Conclusion}
\label{sec:conclusions}

In summary, we have examined the conceptual foundations of continuous pedestrian dynamics models. Starting from a broad context, we have argued that the articulation between the tactical and operational levels of description, which tends to  coincide with the articulation between global path planning and local navigation, raises practical issues for modeling.
While it is generally operated by defining intermediate way-points, the storage in memory of a `tactical' floor field covering all space presents several advantages, particularly in the presence of obstacles or uncomfortable areas on the preferred paths.
Our investigation has focused on
three major branches of models, here dubbed purely reactive, anticipatory, and game-like; it has shed light on the simplifying assumptions under which a branch reduces to another one: differing in their predictions of the future, reactive agents, anticipatory agents, and game players extrapolate future trajectories in spacetime in the form of time-invariant cylinders, cylinders, and flexible tubes, respectively.

For illustrative purposes, an archetypal example was chosen within each modeling branch: the circular specification of the SFM, the ANDA model, and a game in which agents interact via a distance-based repulsive potential. While the first one struggles to replicate head-on collision avoidance at various walking speeds, the latter two produce fairly similar collision-avoiding trajectories. Moreover, the distinction between cognitive processes and mechanical contacts was underscored, at odds with the frequent amalgamation of the two notions in existing models. The effect of the timescales associated with these processes on collision avoidance was studied numerically; the trends predicted by ANDA when the reaction time is increased are qualitatively similar to those reported experimentally in \cite{murakami2022spontaneous}, opening the door to numerical studies of the crowd dynamics of people distracted by their smartphones. The topic is vested with special interest for pedestrian safety in an overly connected society, where accidents due to smartphone-walking are on the surge.

Finally, the insight gained into the relation between the equation of motion of anticipatory models and generalized costs in games proved helpful to naturally account for courtesy or priority effects in the former, as illustrated numerically by a collision avoidance maneuver involving a courteous agent and a standard one and in which the detour is mostly undertaken by the former one.

\rev{However, there is no denying that the proposed framework does not capture all subtleties of digital distraction or courtesy and that even the currently available experimental data only partly uncover the phenomenology that arises in presence of these complex effects. All in all,}
 the development of theoretically better grounded models is strongly advisable when it comes to exploring emerging situations for which one cannot fully rely on the (still scarce) data at hand.\\

\begin{acknowledgments}
We thank Matteo BUTANO for his particularly useful help with game theory. We acknowledge partial financial support by the F{\'e}d{\'e}ration d'Informatique de Lyon (CROSS project) and by the French National Research Agency (Agence Nationale de la Recherche, grant number ANR-20-CE92-0033) and the German Research Foundation (Deutsche Forschungsgemeinschaft DFG, grant number 446168800), in the frame of
the French-German research project MADRAS. 
I. E. acknowledges Ministerio de Economía Competitividad (Spanish Government) through Project No. PID2020-114839GB-I00 MINECO/AEI/FEDER, UE and Asociación de Amigos de la Universidad de Navarra for their economical support. A. R. acknowledges financial support from \'Ecole Normale Sup\'erieure Paris-Saclay.
\end{acknowledgments}

%%% SUPPL MAT

\clearpage
\section*{Supplementary Material}
\subsection*{Integration of a floor field in the SFM.}

To integrate a floor field within the SFM, the desired direction of motion is determined by first computing the gradient of the floor field $\boldsymbol{\phi(r)}$ at the pedestrian’s current position, which yields the \emph{desired} velocity field, $\boldsymbol{v}^{\infty}(\boldsymbol{r})$. The floor field $\phi$ is identical to that pre-computed for the ANDA model; it encapsulates the environmental cost or potential landscape. At each spatial location, the gradient is evaluated via nearest-neighbor interpolation, and the desired velocity is subsequently computed as

\[
\mathbf{v}^{\rm des} (\boldsymbol{r}) = v_{\rm des}\,\frac{-\nabla \phi (\boldsymbol{r})}{\|\nabla \phi (\boldsymbol{r})\|},
\]
where $v_{\rm des}$ denotes the target speed and the negative sign directs the pedestrian towards the steepest descent of the cost field.

Subsequently, the desired force is computed according to
\[
\mathbf{F}_{\rm des}= m\,\frac{\mathbf{v}^{\rm des}(\boldsymbol{r}) - \mathbf{v}}{\tau},
\]
with $m$ representing the pedestrian’s mass, $\mathbf{v}$ his/her current velocity, and $\tau$ a relaxation time constant. We used the same parameter values as for the \emph{naive} SFM (see the main text).

Applying this \emph{hybrid} framework within the obstacle collision avoidance analysis (Section IV-A, main text) enables the agent to navegate through diverse obstacle configurations with minimal perturbation. The resulting trajectories demonstrate that the integration of the floor field significantly enhances collision-free navigation, as evidenced in Fig. S1.

\clearpage
\subsection*{Elliptical Repulsive Force in the SFM.}

In the traditional \emph{circular} specification of the SFM, the long-range repulsive force between agents depends solely on their relative distance. In contrast, the \emph{elliptical} specification updates this interaction by incorporating the anticipated displacement resulting from their relative velocities. In this regard, note that our formulation slightly differs from the original one in Ref.~\cite{helbing1995social}, which relies on an absolute velocity, and follows that implemented e.g. in the UMANS software developed at INRIA \cite{van2020generalized}. The repulsive interaction between pedestrians $i$ and $j$ is given by:
\[
\mathbf{F}_{ij} = A \exp\left(-\frac{b_{ij}}{B}\right) \frac{\|\mathbf{r}_{ij}\|+\|\mathbf{r}_{ij}-\mathbf{y}_{ij}\|}{2b_{ij}} \cdot \frac{1}{2}\left(\frac{\mathbf{r}_{ij}}{\|\mathbf{r}_{ij}\|} + \frac{\mathbf{r}_{ij}-\mathbf{y}_{ij}}{\|\mathbf{r}_{ij}-\mathbf{y}_{ij}\|}\right).
\]
Here, $\mathbf{r}_{ij} = \mathbf{r}_i - \mathbf{r}_j$ is the relative position vector from pedestrian $j$ to $i$, and $\mathbf{y}_{ij} = \Delta t\,(\mathbf{v}_j - \mathbf{v}_i)$ represents the anticipated displacement over a short time horizon $\Delta t$. The effective distance,
\[
b_{ij} = \frac{1}{2}\sqrt{\Big(\|\mathbf{r}_{ij}\| + \|\mathbf{r}_{ij}-\mathbf{y}_{ij}\|\Big)^2 - \|\mathbf{y}_{ij}\|^2},
\]
captures both the current separation and the expected motion. Constants $A$ and $B$ set the force intensity and decay range, respectively.  In our analysis, we have used the same values for $A$ and $B$ as in the main text, and we set $\Delta t=1.5$ seconds \cite{garcia2023limited} (which is a relatively large value). This formulation thus combines instantaneous positions and predicted displacements to yield a directionally averaged force that mitigates potential collisions.

The integration of the \emph{elliptical} formulation within the collision avoidance analysis (Section IV-B, main text) significantly improves the realism of the collision avoidance predicated by the SFM, as substantiated by Fig.~S2.

\clearpage
\subsection*{Supplementary Figures}
\vspace*{\fill}

\begin{figure}[!h]
\centering
\includegraphics[width=0.6\textwidth]{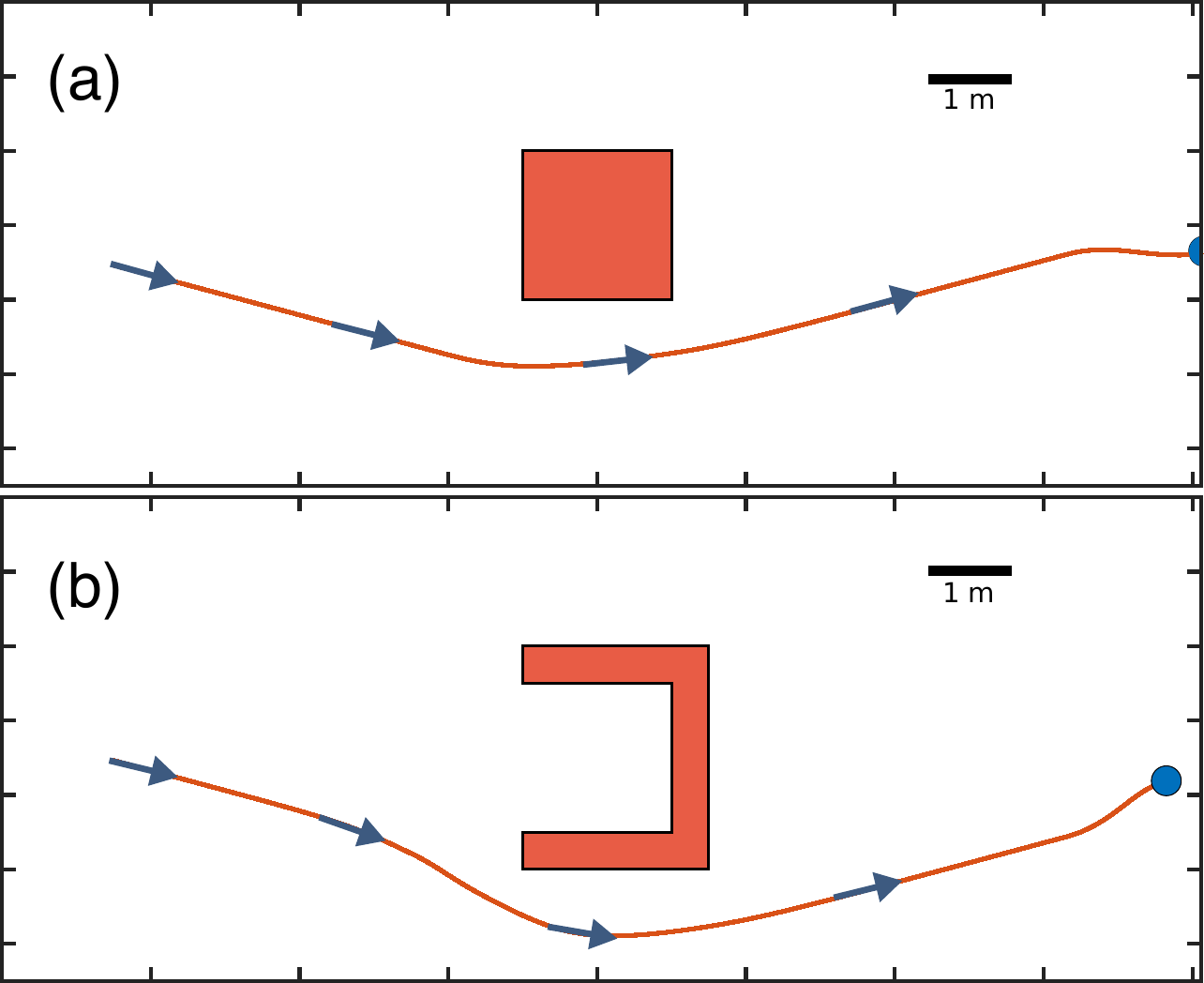}
\caption{Avoidance of a static obstacle simulated with the \emph{hybrid} SFM model complemented with a floor field identical to that of the ANDA model. More precisely, a desired velocity field $\boldsymbol{v}^{\infty}(\boldsymbol{r})$ was computed using the ANDA floor field and used in the SFM.}
\label{fig:SFM_FloorField}
\end{figure}
\vspace*{\fill}

\clearpage
\vspace*{\fill}
\begin{figure}[!h]
\centering
\includegraphics[width=0.9\textwidth]{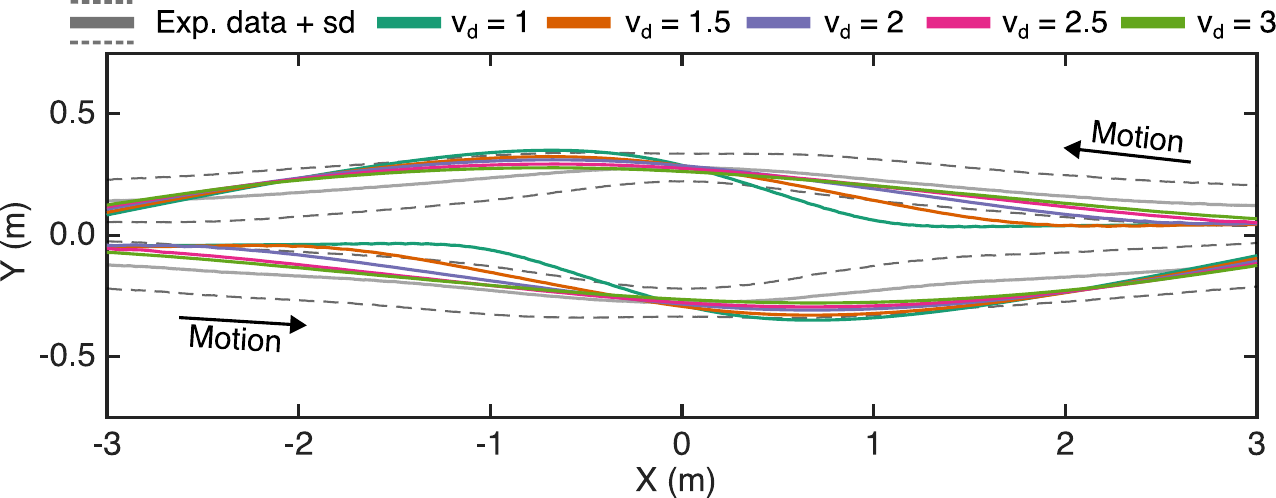}
\caption{Frontal collision avoidance simulated using the \emph{elliptical} specification of the SFM.}
\label{fig:SFM_Ellip}
\end{figure}
\vspace*{\fill}

\end{document}